\documentclass[11pt,twoside]{article}

\usepackage{asp2006}
\usepackage{epsf}
\usepackage{graphicx}
\usepackage{lscape}

\def\Bz{\hbox{$\langle B_z\rangle$}}
\def\gtrsim{\mathrel{\hbox{\rlap{\hbox{\lower4pt\hbox{$\sim$}}}\hbox{$>$}}}}
\def\ltsim{\mathrel{\hbox{\rlap{\hbox{\lower4pt\hbox{$\sim$}}}\hbox{$<$}}}}

\begin{document}

\title{Magnetism of Herbig Ae/Be stars}   
\author{G.A. Wade, E. Alecian, J. Grunhut}   
\affil{Royal Military College of Canada}
\author{C. Catala}
\affil{Observatoire de Paris (LESIA)}    
\author{S. Bagnulo, C.P. Folsom}
\affil{Armagh Observatory}
\author{J.D. Landstreet}
\affil{University of Western Ontario}

\begin{abstract} 
Observations of magnetic fields of stars at the pre-main sequence phase can provide important new insights into the complex physics of the late stages of star formation. This is especially true at intermediate stellar masses, where magnetic fields are strong and globally organised, and therefore most amenable to direct study. Recent circularly-polarised spectroscopic observations of pre-main sequence Herbig Ae/Be stars have revealed the presence of organised magnetic fields in the photospheres of a small fraction of these objects. To date, 9 magnetic HAeBe stars have been detected, and those detections confirmed by repeated observations. The morphology and variability of their Stokes $V$ signatures indicates that their magnetic fields have important dipole components of $\sim$kG strength, and that the dipole is stable on timescales ofat least years. These magnetic stars exhibit a large range of stellar mass, from $\sim 2-13~M_\odot$, and diverse rotational properties, with $v\sin i$ from a few km/s to $\sim 200$~km/s. Most magnetic HAeBe stars show approximately solar abundances; they clearly do not generally exhibit the strong and systematic peculiarities of the magnetic main sequence A and B type stars (the Ap/Bp stars). The observed fractional bulk incidence of magnetic HAeBe stars is about 7\%, a value compatible with the incidence of magnetic intermediate-mass stars on the main sequence. This low incidence is at odds with formation scenarios generally involving magnetically-mediated accretion. The similarily between the magnetic properties of the pre-main sequence and main sequence intermediate-mass stars appears compatible with the hypothesis of a fossil origin of magnetism in these objects.\end{abstract}



\section{Introduction}

Herbig Ae/Be (HAeBe) stars are pre-main sequence (PMS) stars of intermediate mass (Herbig 1960; Hillenbrand et al. 1992), characterised by spectral types
A and B with strong emission lines. They often exhibit infrared excess and are frequently
located within dust-obscured regions and associated with nebulae
(Waters \& Waelkens 1998). According to stellar evolution theory, HAeBe stars should not posses {deep} outer convection zones {which generate the important quantities of outward-flowing mechanical energy required to power an MHD dynamo.} Rather, these stars are expected to have convective cores surrounded by {primarily} radiative sub-photospheric envelopes (Iben 1965; Gilliland 1986). However, since 1980, repeated observations (e.g. Praderie et al. 1982; Catala et al. 1986;
Hamann \& Persson 1992; Pogodin et al. 2005) have shown that many HAeBe stars are intensely active. In particular, some stars display characteristics often associated with magnetic activity and the presence of chromospheres or coronae (e.g. Skinner \& Yamauchi 1996). These properties {have been proposed as indicators that} many of these stars {or their circumstellar envelopes} are intensely magnetically active. 

This proposal has important implications for our picture of how intermediate-mass stars form. In lower-mass pre-main sequence T Tauri stars, it is now generally supposed that accretion is mediated by the presence of strong, large-scale magnetic fields (e.g. discussion by Johns-Krull et al. (1999) and references therein). Some authors (e.g. Muzzerole et al. 2004) have suggested that similar ``magnetospheric accretion'' may occur in intermediate-mass PMS stars as well. 

The magnetic properties of the well-studied main sequence Ap and Bp stars provide a useful context for discussion of the magnetism in HAeBe stars. A few percent (e.g. Wolff 1968, Auri\`ere et al. 2007, Power et al. 2008) of all A and B type {main sequence stars} exhibit organised magnetic fields with strengths ranging from a few
hundred to a few tens of thousands of gauss (e.g. Borra \& Landstreet 1980; Mathys et al. 1997). Because, like their pre-main sequence progenitors, main sequence A and B type stars have radiative envelopes, these magnetic fields are not believed to result from a contemporaneous MHD dynamo. Rather, they are thought to be {\em fossil fields}: the passively-decaying remnants of magnetic fields produced at an earlier convective evolutionary stage, or swept up during star formation. The presence of these fields has important consequences for the structure of
the atmospheres of these Ap/Bp stars, suppressing {large-scale mixing} and
leading to strong atmospheric chemical peculiarities and abundance patches (e.g. Folsom et al. 2006; Lueftinger et al. 2003). Although the separation and mixing processes leading to these chemical peculiarities and abundance structures are not well understood, the peculiarities are sufficiently unique to the Ap/Bp stars that they allow for the robust identification of magnetic stars from high- and even moderate-resolution spectroscopy (i.e. without any direct detection of the magnetic field; Auri\`ere et al. 2007). The presence of the field also appears to strongly influence stellar rotation, with Ap/Bp stars rotating substantially more slowly than other stars of the same spectral type (e.g. Abt \& Morrell 1995). As proposed by St\c{e}pie\'n (2000), the slow rotation of Ap/Bp stars can be explained by angular momentum loss at the pre-main sequence phase, taking into account accretion of matter along the magnetic field lines, the stellar field-disc interaction and a magnetised stellar wind.

Observations of magnetic fields in Herbig Ae/Be stars can therefore serve to address several important astrophysical problems: (1) The role of magnetic field in mediating accretion, and the validity of models which propose that HAeBe stars can be viewed as higher-mass analogues of the T Tau stars. (2) The origin and properties of the magnetic fields of main sequence A and B type stars. (3) The development and evolution of chemical peculiarities and chemical abundance structures in the atmospheres of A and B type stars. (4) The loss of rotational angular momentum which leads to the slow rotation observed in main sequence A and B type stars.

This paper will review recent magnetic observations of Herbig Ae/Be stars, and will discuss how these new data help to clarify the issues listed above.

\section{Searches for magnetic fields in Herbig Ae/Be stars}

During the past 25 years several teams have investigated the magnetic properties of Herbig Ae/Be stars. The general approach has been to search for fields in the atmospheres or envelopes of these stars using circular polarisation spectroscopy to measure the longitudinal Zeeman effect in spectral lines (e.g. Landstreet et al., these proceedings). Catala et al. (1993) attempted to detect Zeeman circular polarisation in the Fe~{\sc ii} $\lambda$5018 and He~{\sc i}~$\lambda 5876$ lines in the spectrum of the prototypical HAeBe star AB Aur, with no detection of a magnetic field, and with upper limits on the order of 1 kG. A survey undertaken by Glagolevski \& Chountonov (1998), including a larger sample of stars, also found no fields, but with relatively poor precision. Ultimately, Donati et al. (1997) and Donati (2000) detected and confirmed the presence of a circular polarisation signature in metallic lines of the Herbig Ae star HD~104237, providing the first detection of a magnetic field in a HAeBe star. 

During the past $\sim 5$ years, research in this field has been intense. The remainder of this Section is devoted to these more recent investigations, undertaken primarily using the low-resolution FORS1 instrument at the ESO-VLT, and the high-resolution spectropolarimeter ESPaDOnS at the Canada-France-Hawaii Telescope (CFHT).

\subsection{Low-resolution spectropolarimetry with FORS1}

Wade et al. (2007) presented results of a survey of the longitudinal magnetic fields of HAeBe stars conducted using the low-resolution FORS1 spectropolarimeter at the VLT observatory. Comprising 68 observations of 49 HAeBe stars, this work is the largest and most sensitive search for magnetic fields in HAeBe stars published to date. 

\begin{figure*}
\includegraphics[width=14cm]{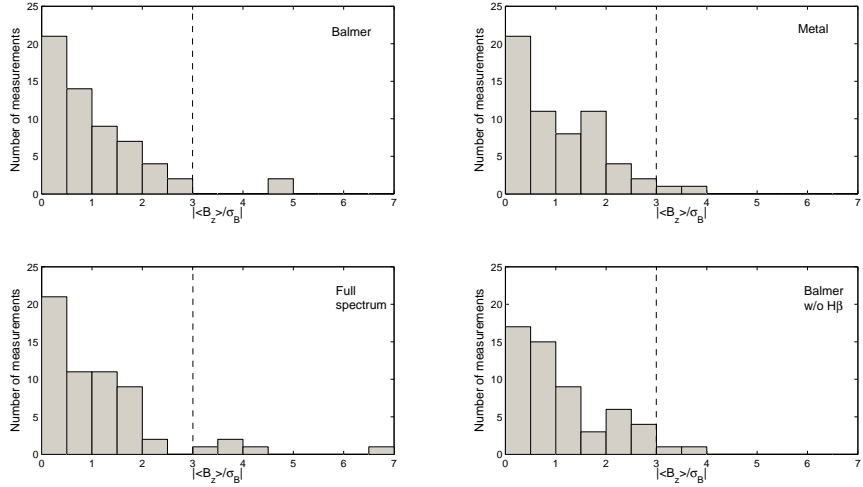}
\caption{Histograms of the detection significance $z=\Bz/\sigma_B$ of magnetic field measurements of stars observed by Wade et al. (2007). Each panel corresponds to one of the 4 measurements obtained using different parts of the spectrum. The dashed line indicates the $3\sigma$ detection threshold. Nearly all measurements correspond to non-detections, and formal detections are obtained for only 4 stars.}\label{zBz}
\end{figure*}

The primary goal of this survey was to determine whether strong, organised magnetic fields, similar to those of main sequence Ap/Bp stars, could be detected in the HAeBe stars, and if so, to determine the intensities, geometrical characteristics, and statistical incidence of such fields.

With this aim, Wade et al. (2007) obtained high-S/N ($\sim 3000$ per \AA), low resolution ($R\simeq 1000-1500$) spectral ($\lambda=3450-5900$~\AA) observations of HAeBe stars in circularly polarised light. The data were reduced and analysed using the procedures of Bagnulo et al. (2006). The longitudinal magnetic field was inferred from the polarisation spectra using a linear least-squares fit based on the predicted circularly polarised flux $V/I$ in the weak-field regime. This procedure was repeated 4 times for each spectrum, diagnosing the field from all Balmer lines, all Balmer lines except H$\beta$, metallic lines, and the full spectrum. The mean 1$\sigma$ longitudinal field uncertainty measured from Balmer lines was 66~G, while that measured using the full spectrum was 48~G. 
 
The histograms of the detection significance $z=|\Bz/\sigma_B|$ of the resultant measurements are shown in Fig. 1. To robustly treat cases for which the longitudinal field was detected at about the 3$\sigma$ level, and in which minor changes in the data reduction would transform a marginal detection in a null or into a definite detection, Wade et al. (2007) evaluated the consistency of the longitudinal fields determined from the various spectral regions using two detection criteria described by Bagnulo et al. (2006). This allowed them to realistically judge the true significance of apparent detections.

Ultimately, Wade et al. (2007) concluded tentatively that weak longitudinal magnetic fields were probably detected in spectra of two HAeBe stars: HD 101412 (obtaining a measurement of $512\pm 111$~G from Balmer lines) and BF Ori (obtaining a measurements of $-180\pm 38$~G from Balmer lines), and possibly detected in two other stars: HD~36112 and CPD-53295.

Magnetic observations of a small number HAeBe stars (about 14) have also been carried out using FORS1 by Hubrig et al. (2004, 2006, 2007), who reported marginal detections of magnetic fields in several stars. Wade et al. (2007) observed meny of the stars reported to be magnetic by these authors, and obtained no detection of magnetic field in any of them. {Hubrig et al. (2007) also reported the presence of strong Stokes $V$ signatures in spectral lines of some HAeBe stars, which they interpreted to be due to ``circumstellar'' magnetic fields. Wade et al. (2007) were unable to confirm the detection of such features in their own spectra. As a further test, to check if they could confirm the strong polarisation in the same data employed by Hubrig et al., they extracted from the ESO Science Archive 3 spectra of the HAeBe star HD 190073 discussed by Hubrig et al. (2007), reducing and analysing them according to the procedures of Bagnulo et al. (2006). Remarkably, Wade et al. found no evidence for the presence of strong, systematic polarisation signatures reported by Hubrig et al. (2007). 

\subsection{High-resolution spectropolarimetry with ESPaDOnS}


Alecian, Catala, Wade and collaborators have also undertaken large circular polarisation surveys of HAeBe stars using the high-resolution ESPaDOnS spectropolarimeter at the Canada-France-Hawaii Telescope, as well as other high-resolution instruments. Although the surveys have been completed, the general results have not yet been published in refereed journals. However, an overview is reported by Alecian et al. (2008a), who describe a "field" survey of brighter HAeBe stars (aimed at determining the general magnetic properties of these stars), and a "cluster" survey (aimed at investigating the evolution of magnetism and rotation). In addition, results for individual stars of great interest have been reported by Wade et al. (2005), Catala et al. (2007a), Folsom et al. (2008) and Alecian et al. (2008b, 2008c).


\begin{figure*}[!b]
\includegraphics[width=5cm,angle=-90,viewport=50 0 320 750,clip]{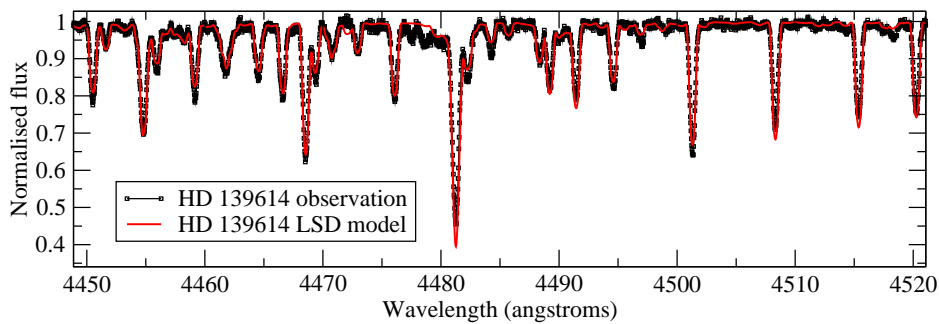}
\caption{A small region of the high-resolution ESPaDOnS spectrum of the HAeBe star HD~139614 (black squares), compared to a solar abundance, LTE photospheric synthetic spectrum (red curve) corresponding to the line mask employed to extract LSD profiles of this star by Wade et al. (2005). The agreement between the observations and the theoretical model is excellent, indicating that the LSD approach is well-suited to HD~139614 and stars with analogous spectral properties.}
\end{figure*}

The ESPaDOnS surveys comprise over 200 circular polarisation spectra of about 130 HAeBe stars, including many of the stars observed by Wade et al. (2007) and Hubrig et al. (2004, 2006, 2007). The spectra cover the range 3800-10400~\AA\ with a resolving power $R\simeq 65000$, and a median peak S/N of about 1450:1 per \AA\ at 5000~\AA. This is about one-half the mean S/N of the FORS1 spectra of Wade et al. (2007); it furthermore varies more strongly with wavelength, decreasing to about 1/2 its peak value at 4000~\AA\ and 9000~\AA. However, the excellent results obtained with ESPaDOnS so far (see e.g. Donati et al. 2005, 2006ab; Catala et al. 2007a, 2007b; Landstreet et al. 2008; Auri\`ere et al. 2008, Petit et al. 2008) demonstrate that the advantage of substantially higher resolving power (45-65 times better resolution) easily compensates for this difference. In particular, the ESPaDOnS measurements of HAeBe stars are characterised by a substantially higher magnetic field detection and confirmation rate, more robust detections of weak longitudinal fields, straightforward confirmation of field detections, a lower ultimate detection threshold, and sensitivity to more complex field topologies. In addition, ESPaDOnS and Narval data provide high-resolution line profiles, allowing detailed studies of the circumstellar environment, photospheric properties and structures, chemical abundances and stellar rotation (see e.g. Fig. 2).

The ESPaDOnS observations are typically analysed using the Least-Squares Deconvolution (LSD; Donati et al. 1997) multi-line procedure, extracting polarisation information from hundreds or thousands of lines in the observed spectrum to obtain the magnetic field diagnosis. The resultant high-S/N mean profiles allow the diagnosis of the magnetic field directly from the presence (or absence) of resolved Stokes $V$ signatures within essentially all stellar spectral lines (see Fig. 3).  The longitudinal field can also be diagnosed: it is inferred from the first moment of the Stokes $V$ LSD profile (e.g. Wade et al. 2000). However, because ESPaDOnS is a high-resolution spectropolarimeter, the inferred errors are very sensitive to the line profile characteristics - in particular $v\sin i$ and emission contamination. This results in a substantially broader distribution of longitudinal field uncertainties, the largest of which (for stars with the highest $v\sin i$ or lines most contaminated by emission) reach several hundred gauss. However, for the majority of the sample, the longitudinal field uncertainties are competitive with those obtained using FORS1, and for those stars which are most suitable spectroscopically, the longitudinal field uncertainties are often better than those from FORS1.

\begin{figure*}[!t]
\includegraphics[width=6.5cm,angle=90]{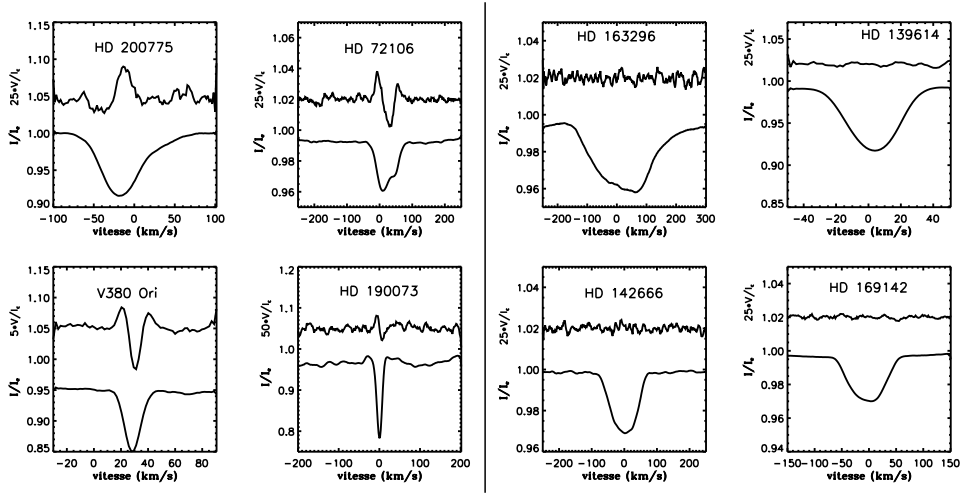}
\caption{Least-Squares Deconvolved (LSD) Stokes $I$ and $V$ profiles of Herbig stars observed in the context of the ESPaDOnS survey in which magnetic fields are detected (left), and those in which magnetic fields are not detected (right).}
\end{figure*}

Similar to the FORS1 survey, null magnetic field results are obtained for the majority of stars observed, including HD 144432, HD 31648 and HD 139614, in which Hubrig et al. (2004, 2007) have claimed marginal magnetic field detections, and in HD~36112 and BF~Ori, for which FORS1 data were reported to be "suggestive" of magnetic fields (Wade et al. 2007). Nor do the ESPaDOnS spectra yield evidence for the systematic presence of polarisation signatures in lines diagnostic of the circumstellar environment, such as Ca~{\sc ii} H and K, Mg~{\sc ii} $\lambda 4481$, the H$\alpha$ emission line, the O~{\sc i} $\lambda\lambda$6300, 6364 emission lines, the IR calcium triplet, etc.  

On the other hand, magnetic fields are clearly detected in photospheric metal lines of a small number of the stars observed with ESPaDOnS: HD~190073 (Catala et al. 2007a), HD~72106 (Folsom et al. 2008), HD~200775 (Alecian et al. 2008b), V380 Ori (Wade et al. 2005, Alecian et al. 2008a), NGC 6611-601 and NGC 2244-OI 201 (Alecian et al. 2008c). Analogous observations have also been acquired using UCLES+Semelpol at the Anglo-Australian Telescope by S. Marsden which confirm the detection of fields in HD~104237 and HD 101412 (Wade et al., in prep.). All of these stars have now been detected in multiple observations during different observing runs, and three of these stars have been shown to exhibit coherent, periodic variations of their Stokes $V$ profiles (Alecian et al. 2008a, 2008b, Folsom et al. 2008). LSD profiles illustrating the detection of fields in some these stars are shown in Figs. 3 and 4. Fig. 3 also contrasts the LSD profiles of the detected stars to those of a number of undetected stars.

Hubrig et al. (2007) claim that LSD is inappropriate for magnetic field diagnosis of HAeBe stars due to the complexity of their spectra, and due to contributions to the metallic line spectrum from the circumstellar environment. This claim is refuted in Fig. 2, for the particular case of HD~139614, where the observed spectrum is shown to be almost perfectly reproduced by the LSD model - as well as essentially any main sequence star. Moreover, the field detections achieved using LSD that are cited above - detections that are repeatable, coherent and interpretable within the framework of simple models - demonstrate that LSD can be used effectively for field diagnosis in HAeBe stars, even those stars with very complex spectra (e.g. the spectra of HD~190073 and V380 Ori, which are tremendously more complex than that of HD~139614). This conclusion is supported by the analysis of the large body of data accumulated within the context of the ESPaDOnS survey. 

\begin{figure}[!b]
\centering
\includegraphics[width=3cm,angle=90]{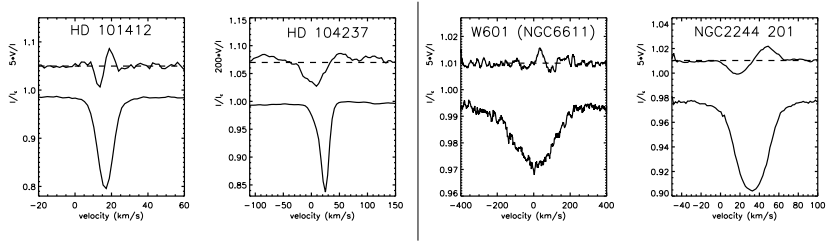}
\caption{Additional LSD profiles of magnetic HAeBe stars observed using high-resolution spectropolarimetry: HD 104237, HD 101412, W601 and NGC~2244~201 (Wade et al. in prep.; Alecian et al. 2008c.).}
\label{fig:autre}
\end{figure}

\section{Properties of magnetic HAeBe stars}

\begin{figure*}[t!]
\centering
\includegraphics[width=8cm]{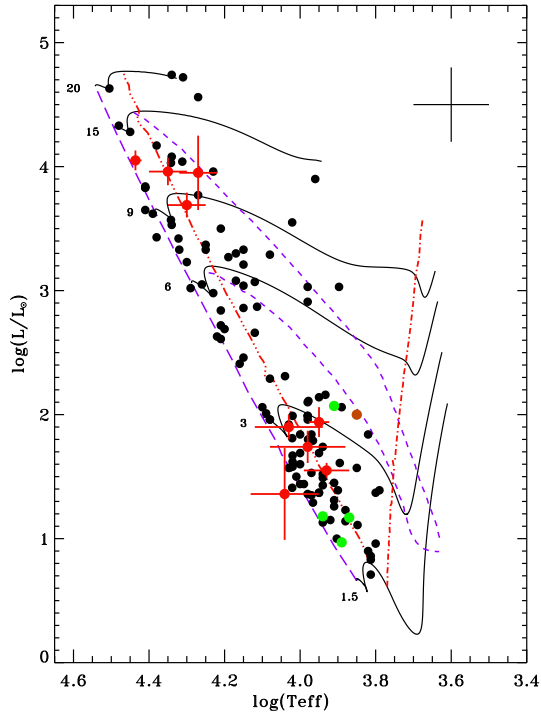}
\caption{HR diagram showing the field and cluster Herbig stars observed by Wade et al. (in prep.). Undetected HAeBe stars (black circles), confirmed magnetic HAeBe stars (red dots with error bars) and unconfirmed detection claims (green and orange circles).The birthlines for 10$^{-5}$ and 10$^{-4}$ $M_{\odot}$\,yr$^{-1}$ mass accretion rate are plotted using blue short-dashed lines (Palla \& Stahler 1993), while the blue long-dashed line is the ZAMS. PMS evolutionary tracks (full black lines), the convective envelope disappearance (red dot-dashed line) and the convective core appearance (red dot-dot-dot-dashed line), calculated with CESAM (Morel 1997) are also plotted.}
\label{fig:hr}
\end{figure*}

The Herbig Ae/Be stars in which magnetic fields are detected correspond to 7\% of the sample observed with ESPaDOnS. This bulk fractional incidence is remarkably similar to the canonical incidence of 5-10\% (e.g. Wolff 1968) of magnetic Ap and Bp stars on the main sequence. The results reported in the papers cited above show that the magnetic fields of all detected HAeBe stars are organised on large scales, and that the fields of the stars that have been studied in detail (V380 Ori, HD~72106A and HD~200775) have important dipole components with characteristic polar strengths of order 1~kG.} For those stars which have been observed many times, the magnetic Stokes $V$ signatures are shown to vary periodically, and to show no long-term evolution. This suggests that the dipole fields are stable on timescales of at least years. These properties are all analogous to those of the main sequence Ap/Bp stars. A more detailed discussion of the magnetic topologies of HAeBe stars is provided by Alecian et al. (these proceedings). 

As shown in Fig. 5, the detected stars span a large range of mass and PMS evolutionary state. Many have spectra that are strongly influenced by the circumstellar environment, and some are members of multiple systems. Apart from a tendancy to slow rotation, the magnetic HAeBe stars are superficially undistinguishable from the general population. In particular, the magnetic stars do not generally exhibit outstanding levels of spectroscopic or photometric activity.

\section{Magnetic constraints on undetected stars}

For the large majority of stars studied by Wade et al. (2007) and in the ESPaDOnS survey, no magnetic fields were detected. Wade et al. (2007) reported the results of Monte Carlo simulations aimed at modeling their FORS1 measurements. They computed synthetic distributions of longitudinal field measurements assuming populations of stars with various dipolar magnetic field characteristics. Comparing the observed and computed distributions, they concluded that their observations had the following properties: they were consistent within statistical uncertainty with a distribution of non-magnetic stars; they were inconsistent with a uniform population of magnetic stars with aligned dipole magnetic fields, if their dipole intensities $B_{\rm d}\gtrsim 300$~G; and they were inconsistent with a uniform population of magnetic stars with perpendicular magnetic fields, if $B_{\rm d}\gtrsim 500$~G. Inclusion of the ESPaDOnS field survey results in this analysis (Wade et al., in prep.) supports these conclusions, and reduces the upper limit somewhat.

\begin{figure*}[!t]
\includegraphics[width=10cm,angle=-90]{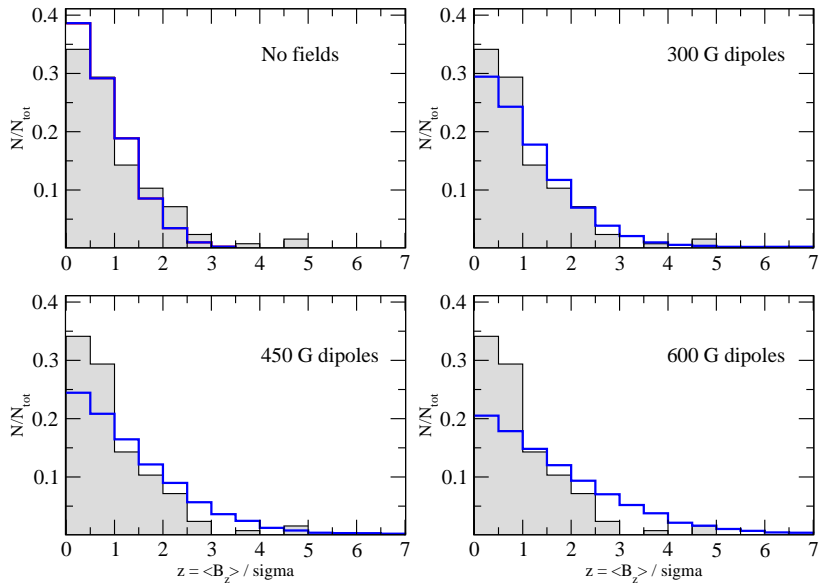}
\caption{Histograms of the detection significance of the longitudinal magnetic field, measured from each of the ESPaDOnS and FORS1 spectra. A one-sided Kolmogorov-Smirnov test was performed on the cumulative distributions of longitudinal fields to quantitatively test whether the observed and computed distributions are representative of the same population. Based on this analysis, Wade et al. (in prep.) are able to rule out uniform populations of stars with dipolar fields above about 300~G.}
\end{figure*}

Fig. 6 compares the observed histogram of uncertainty-normalised longitudinal field measurements $z=\langle B_z\rangle/\sigma_B$ (including both FORS1 and ESPaDOnS data), with synthetic histograms compiled from the Monte Carlo simultations.\footnote{We underscore that the inferred field intensities refer to the dipole field polar strength at the stellar surface, and not to the mean longitudinal field. }




Wade et al. (in prep; see also Wade et al. 2008) also used the LSD profiles of the undetected Herbig stars to constrain the intensities of undetected magnetic fields.  They employed the same model populations of magnetic stars developed by Wade et al. (2007) to create synthetic Stokes $V$ LSD profiles corresponding to each of the ESPaDOnS observations (using the profile synthesis procedure described by Alecian et al. 2007b), and introduced synthetic Gaussian noise corresponding to the noise level in the real LSD $V$ profile. Finally, for each synthetic LSD profile they evaluated the probability that a Stokes $V$ signature was detected, using the same criteria that are applied in the real LSD procedure (see Donati et al. 1997). Again, this procedure was repeated 100 times for each observation and for each model, using different realisations of the randomly-selected variables. Examples of observed and synthetic LSD profiles are shown in Fig. 7.

Even for models as weak as 300 G, Wade et al. obtain detections of small numbers of stars in over 90\% of model realisations. This result is consistent with that derived from the longitudinal field measurements, and demonstrates that the LSD profiles strongly constrain models which propose the presence of weak, organised magnetic fields in all HAeBe stars.  On the other hand, there may still exist a small number of magnetic stars, with magnetic properties similar to the detected magnetic HAeBe stars, present in the undetected sample. We note however that the dipole intensities derived by Alecian et al. (2008a, 2008b) and Folsom et al. (2008) for the detected magnetic HAeBe stars, when evolved to the main sequence under the assumption of flux conservation (Table 1 of Alecian et al. 2008a), are typical of those of the majority of Ap/Bp stars (Power et al. 2008). This indicates that the Herbig stars in which fields have been detected do not have unusually strong fields in comparison to other intermediate-mass stars, and therefore that most of the magnetic stars in the sample are very probably already detected.


\begin{figure*}
\includegraphics[width=4.25cm,angle=-90]{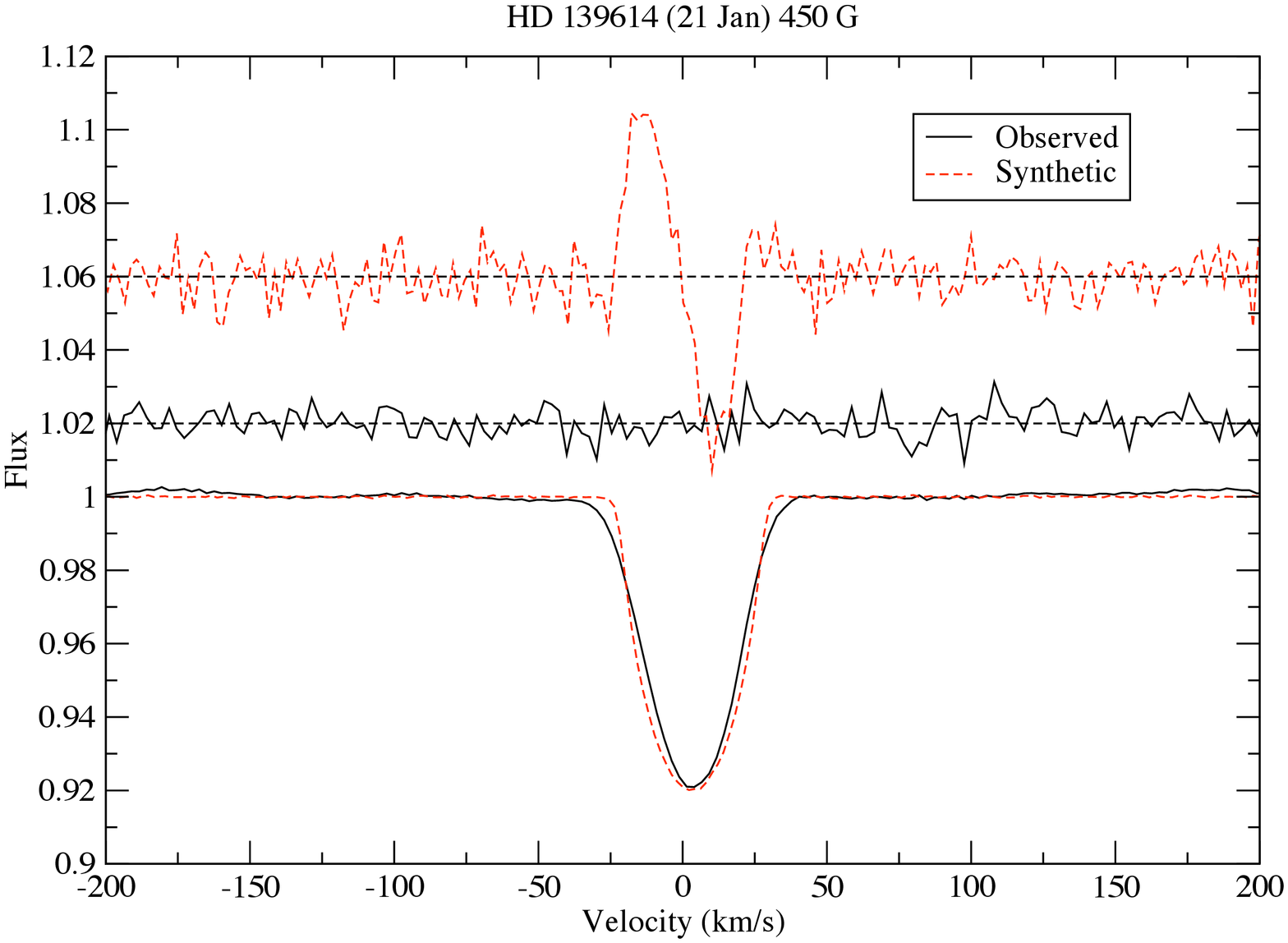}\includegraphics[width=5cm,angle=-90]{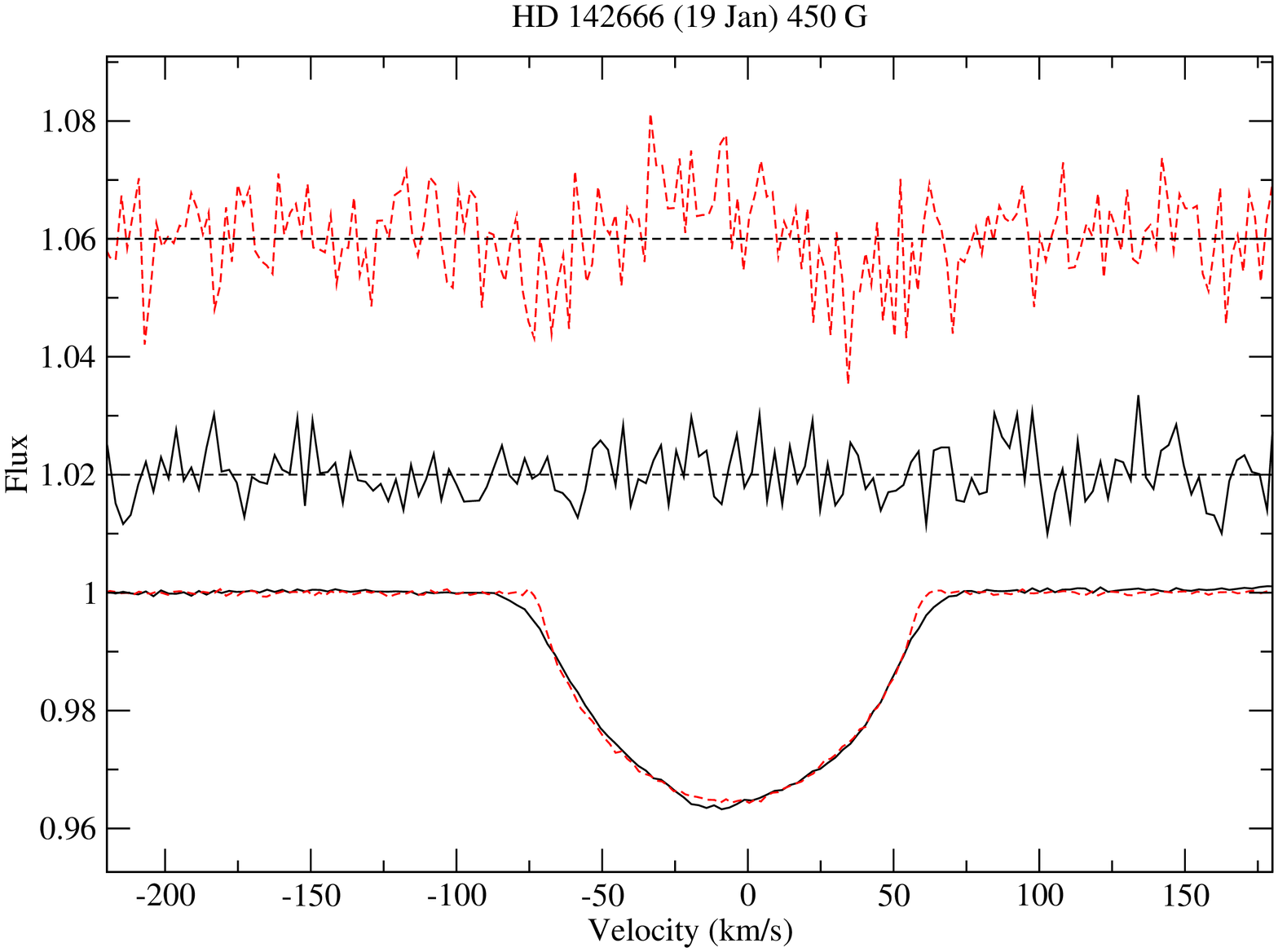}
\caption{Observed and synthetic LSD profiles for the HAeBe stars HD 139614 and HD 142666. Solid (black) curves represent the observations, while dotted (red) lines correspond to the synthetic Stokes $I$ and $V$ profiles for a 450~G dipole field. Both simulations correspond to definite detections, indicating that such fields, if present, would be easily detected in these observations.}
\label{f1}
\end{figure*}

\section{Activity, magnetospheric accretion, chemical peculiarity and rotation}

Some investigators have interpreted the strong spectroscopic, polarimetric and photometric activity of HAeBe stars to indicate that these stars are magnetically active (e.g. Catala et al. 1986). The magnetic results described in this paper do not support the proposal the the activity of most HAeBe stars is of magnetic origin.  {Some investigators have also suggested that HAeBe stars are the higher mass analogues of classical T Tauri stars (CTTS), and that all of the phenomena associated with CTTS are also operating in HAeBe stars, including magnetospheric accretion (see for example Muzerolle et al. 2004).  

Magnetospheric accretion requires the presence of strong, large-scale magnetic fields at the stellar surface, and in the case of CTTS the predicted field strengths range up to several kG for specific stars. Wade et al. (2007) used their data to evaluate three magnetospheric accretion models (see Johns-Krull et al. 1999). Estimating typical values of the mass, radius, rotation period and accretion rate, they computed that a dipole magnetic field with an intensity of about 500~G is required for magnetospheric accretion according to the models of K\"onigl  (1991) and Shu et al. (1994), and of about 100~G for the model of Cameron \& Campbell (1993). Based on the results of their Monte Carlo simulations, they concluded that magnetospheric accretion according to the theories of K\"onigl (1991) and Shu et al. (1994) is not generally occurring in HAeBe stars. Ongoing characterisation of the magnetic strengths and geometries of the detected magnetic HAeBe stars should allow a detailed comparison of their properties with the predictions of all three models, and perhaps with more sophisticated numerical simulations (e.g. Yalenina et al. 2006).

A characteristic observational feature of magnetic A and B type stars on the main sequences is their strong photospheric chemical peculiarity. Although the basic mechanism responsible for the production of this phenomenon is known, the roles of the various separation and mixing processes, and the influence of the magnetic field, are understood only schematically. The new observations of HAeBe stars provide the potential for the study of these phenomena at their earliest stages, allowing us to investigate the conditions and timescales required for the development of chemical peculiarity. Although this aspect has not yet been investigated in much detail, the ESPaDOnS observations suggest that chemical peculiarity is detected in only two of the magnetic stars identified so far: HD~72106A and NGC 6611-601. As is reported by Folsom et al. (2008), HD~72106 shows both strong peculiarities analogous to those of the Ap/Bp stars, as well as line profile variability indicative of chemical abundance patches. Alecian et al. (2008c) claim that NGC 6611-601 may be a PMS helium-strong star.

As discussed earlier, magnetic A and B type stars on the main sequence rotate significantly more slowly than non-magnetic stars of the same spectral type. This suggests that the magnetic field plays an important role in the shedding of rotational angular momentum in the magnetic stars, probably at the pre-main sequence stage (e.g., see St\c{e}pie\'n 2000). Alecian et al. (in prep) explore this aspect in some detail, first by using their measured $v\sin i$s, and more accurately by using their directly-inferred rotational periods and radii. Many of the detected stars appear to be slow rotators - hence significant angular momentum already seems to have been shed by these magnetic stars, even at these relatively young ages. On the other hand, some of these stars are rapid rotators (e.g. Alecian et al. 2008b, Folsom et al. 2008).

\section{Conclusion}

{In this paper we have reviewed the results of recent large-scale surveys of the magnetic properties of Herbig Ae/Be stars, demonstrating the clear detection of strong, organised magnetic fields in a small fraction of the observed stars, and the apparent absence of magnetic fields in the large majority. We have reviewed the observed characteristics of these fields, the physical properties of the detected magnetic HAeBe stars, and the constraints on the absence of fields in the undetected stars. Finally, we have discussed the implications of these new data for our understanding of activity and magnetospheric accretion in PMS intermediate-mass stars, for the evolution of rotational angular momentum, and for the development of photospheric chemical peculiarity.

\acknowledgements 
GAW and JDL acknowledge support from the Natural Science and Engineering Research Council of Canada (NSERC). GAW acknowledges support from the Academic Research Programme of the Department of National Defence (Canada).



\end{document}